\documentclass[aps,twocolumn,superscriptaddress]{revtex4}
\usepackage{graphicx}
\usepackage{color}
\usepackage[normalem]{ulem}

\begin{document}

%\preprint{}

\title{%%Low-energy $\mu$SR as a novel local probe technique to study photo induced effects
%%on a nm scale\\ or\\
%Local probe studies of photo induced effects on a nm scale \\or \\
%Local probe studies of photogenerated charge carriers on a nm scale in Si and Ge
%Photo-induced inversion of germanium and silicon in a 200-nm-deep surface region}
Photo-induced persistent inversion of germanium in a 200-nm-deep surface region}


\author{T.~Prokscha}
\email[Correspondance and requests for materials should be addressed to T.P.]{(thomas.prokscha@psi.ch)}
\affiliation{Paul Scherrer Institute, Laboratory for Muon Spin Spectroscopy, CH-5232 Villigen PSI, Switzerland}
\author{K.H.~Chow}
\affiliation{Department of Physics, University of Alberta, Edmonton T6G 2E1, Canada}
\author{E.~Stilp}
\affiliation{Paul Scherrer Institute, Laboratory for Muon Spin Spectroscopy, CH-5232 Villigen PSI, Switzerland}
\affiliation{Physics Institute, University of Zurich, 8057 Zurich, Switzerland}
\author{A.~Suter}
\affiliation{Paul Scherrer Institute, Laboratory for Muon Spin Spectroscopy, CH-5232 Villigen PSI, Switzerland}
\author{H.~Luetkens}
\affiliation{Paul Scherrer Institute, Laboratory for Muon Spin Spectroscopy, CH-5232 Villigen PSI, Switzerland}
\author{E.~Morenzoni}
\affiliation{Paul Scherrer Institute, Laboratory for Muon Spin Spectroscopy, CH-5232 Villigen PSI, Switzerland}
\author{G.J.~Nieuwenhuys}
\affiliation{Paul Scherrer Institute, Laboratory for Muon Spin Spectroscopy, CH-5232 Villigen PSI, Switzerland}
\affiliation{Kamerlingh Onnes Laboratory, Leiden University, 2300 RA Leiden, The Netherlands}
\author{Z.~Salman}
\affiliation{Paul Scherrer Institute, Laboratory for Muon Spin Spectroscopy, CH-5232 Villigen PSI, Switzerland}
\author{R.~Scheuermann}
\affiliation{Paul Scherrer Institute, Laboratory for Muon Spin Spectroscopy, CH-5232 Villigen PSI, Switzerland}

\date{\today}

% \begin{abstract}
%  Photogenerated charge carriers. Low-energy moun spin rotation (LE-$\mu$SR).
% \end{abstract}

%\pacs{78.70.-g 72.20.Jv 76.75.+i}
%\keywords{}

\maketitle

%
%------------------- body of paper begins here -------------------------------
%
%
\textbf{
The controlled manipulation of the charge carrier concentration in nanometer thin
layers is the basis of current semiconductor technology and of fundamental
importance for device applications. Here we show that it is possible to induce a 
persistent inversion from n- to p-type in a 200-nm-thick surface layer of a 
germanium wafer by illumination with white and blue light. We induce the inversion with a
half-life of $\sim$~12 hours at a temperature of 220~K which disappears above 280~K.
The photo-induced inversion is absent for a sample with a 20-nm-thick gold capping
layer providing a Schottky barrier at the interface. This indicates that charge accumulation
at the surface is essential to explain the observed inversion.
The contactless change of carrier concentration is potentially interesting
for device applications in opto-electronics where the
gate electrode and gate oxide could be replaced by the semiconductor surface.
}

%%%% main text begins here
% \section*{Introduction}
\vspace{10pt}
In photo-conductors, photo-diodes and photo-transistors illumination causes
the generation of free charge carriers in normally depleted regions of the
device \cite{sze2007}.
% For example, in a n-p-n bipolar photo-transistor the base
% is hole doped (p-type) and can be exposed to light. For a floating base the n-type
% collector is positively biased with respect to the n-type emitter. Holes, photo-generated
% in the base-collector depletion region, are trapped in the base and lowering
% the energy barrier for electrons. This allows a large electron flow from the
% emitter to the collector. Photo-generated electrons in the base are drifting
% either to the collector or emitter.
The working principle in these devices is different from %the commonly used
a MOSFET (metal-oxide-semiconductor field-effect-transistor) where
an applied electric field at the gate electrode causes inversion from the
p-type gate to n-type in order to switch the transistor
into a ``current on'' state.
Recent developments on nanowire transistors could offer an alternative to MOSFETs
without the need of layer inversion and junctions \cite{colinge2010}.
Charge coupled devices are another important class of photo sensitive
devices where inversion of a p-type semiconductor is the basis for its functionality:
a gate bias drives the semiconductor substrate to deep depletion and
photo-generated holes are removed from the surface region, while electrons are accumulated
until surface charge saturation is reached. In a range of the depletion width
(up to $\sim$~300~nm) the semiconductor substrate becomes inverted.
In technologically important transparent conducting oxides, such as In$_2$O$_3$
or SnO$_2$, intrinsic or photo-induced charge accumulation at surfaces or interfaces
changes near-surface carrier concentrations, i.e. the electronic properties.
Examples are the n-type conductivity of In$_2$O$_3$ \cite{king_surface_2008}, the
photo-induced Schottky barrier in photorefractive materials with SnO$_2$
electrodes \cite{frejlich_photoinduced_2010}, or the enhancement of photocurrents
in SnO$_2$ nanowires by the decoration with metallic nanoparticles, which is attributed to the formation
of Schottky junctions in the vicinity of the nanoparticles \cite{lin_photocurrent_2008}.

% \begin{figure}[ht]
% \begin{center}
% \includegraphics[width=0.9\linewidth]{LEM_setup}
% \includegraphics[width=1.0\linewidth]{Ge_stopping-light_Profile}
% \end{center}
% \caption{\textbf{Schematic of experimental setup and muon stopping and light absorption profiles in germanium.}
% Muon stopping profiles at the indicated implantation energies were calculated with the Monte-Carlo code {\ttfamily TrimSP}\cite{trimsp,morenzoni_implantation_2002} from 1 keV to 20 keV.
% A light absorption coefficient $\alpha \sim 5\cdot 10^{5}/cm$ has been assumed
% for the used blue light sources with peak wavelengths of 470~nm and 405~nm
% \cite{philipp_optical_1959}, corresponding to photon energies of 2.6 and 3.0~eV, respectively.
% The pressure at the sample was $\leq 10^{-8}$~mbar.}
% \label{fig1}
% \end{figure}

Changes in charge-carrier concentration and dopant type in the bulk are usually detected by
macroscopic measurements (e.g. conductivity, Hall-effect, capacitance-voltage profiling (C-V)) which also require
the presence of junctions and leads. %On the other hand 
Scanning probe microscopy techniques allow to carry out time-resolved 
local mapping of photo-generated charge carriers with lateral resolutions in the sub-hundred nanometer range
and temporal resolution of less than hundred $\mu$s \cite{alexe2012,takihara2008,coffey2006}.
% This is different to the scanning probe techniques where 
Here, the presence of photo-induced charge carriers is detected indirectly by changes of the
resonance frequency of a cantilever or by changes of the potential offset between a cantilever and the sample
surface. By contrast, the positively charged
muon $\mu^+$ represents a very sensitive \textit{local probe} to detect
variations in charge-carrier concentrations as a function of temperature
and doping in semiconductors without the need of electrical contacts or the application of
electric fields/currents\cite{yaouanc2011,patterson_muonium_1988,chow1998,cox2009}. 
In this context \textit{local probe} means that the implanted $\mu^+$, which comes to rest at an
interstitial site, is directly interacting with the free charge carriers in its \textit{local}, nanometer sized 
environment.  

In this paper we demonstrate that we can use illumination to controllably generate persistent charge carriers
beneath the surface of nominally undoped, commercial n-type Ge wafers.
These charge carriers are detected at tunable depths $\lesssim$ 200~nm with
low-energy muon spin rotation (LE-$\mu$SR) \cite{prokscha_new_2008,morenzoni_generation_1994}
using the LE-$\mu$SR spectrometer (LEM) at the Swiss Muon Source S$\mu$S at PSI.
Our results imply that the photo-generated electrons are trapped in surface states which
causes the semiconductor near-surface region to convert from n- to p-type.
This demonstrates that it is in principle possible to create a configuration similar to a
MOSFET without using gate electrodes and gate voltages.

A low-energy $\mu^+$ beam with tunable energies in the keV range allows depth-dependent investigations
of near-surface regions on a nanometer scale, as shown in Fig.~1. 
This is different to conventional \textit{bulk} $\mu$SR studies, where energetic muon beams with an 
energy of typically 4~MeV are used to stop the muons deep inside the sample at depths of the order of 
hundred micrometer.
Note that muon implantation 
energies $<$ 14~keV result in average muon stopping depths that are well-matched with the absorption 
profile of light in Ge.
%
% The implanted $\mu^+$ comes to rest at an interstitial site.
In a semiconductor or insulator it can capture one or two electrons to form
the different charge states of muonium [Mu, ($\mu^+e^-$), mass of $\mu^+ \simeq 1/9$ proton mass].
Depending on factors such as the dopant concentration and Mu formation energy it occurs in
one of three, possibly coexisting, charge states Mu$^+$, Mu$^0$ or Mu$^-$, analogous
to hydrogen\cite{yaouanc2011,patterson_muonium_1988,chow1998,cox2009}.
The muon decays with a lifetime of 2.2~$\mu s$ into a positron and two neutrinos.
The detection of the various Mu states occurs by observation of the anisotropic
distribution of decay positrons which are preferentially emitted along the
$\mu^+$ spin direction. This anisotropic decay is the basis for the muon
spin rotation technique ($\mu$SR) \cite{yaouanc2011}.
In $\mu$SR experiments on
semiconductors, where the muonium has been often used as a light pseudo-isotope
of isolated hydrogen, the different Mu charge states can be distinguished by
their different spin precession signatures in a magnetic field $B$ applied transverse
to the initial muon spin direction. In case of a  Mu$^+$ or a  Mu$^-$ state the
muon spin precesses with a frequency $\omega_{\mu} = \gamma_{\mu} B$,
where $\gamma_{\mu}/2\pi = 135.54$~MHz/T is the gyromagnetic ratio of the muon.
This precession appears as an oscillation in the recorded muon decay histograms of
several positron detectors surrounding the sample.
The amplitude $A$ of this oscillation is the muon decay asymmetry and is proportional to
the polarization of the corresponding Mu state.
In Mu$^0$ the hyperfine coupling
% with the unpaired electron generates an additional effective magnetic field at the muon site
causes the muon spin to precess at frequencies that are much higher than $\omega_{\mu}$.
This allows one to clearly distinguish Mu$^0$ states from Mu$^+$ and Mu$^-$.
The latter are often referred as \textit{diamagnetic} Mu$^D$ which means a $\mu^+$ without
unpaired electrons nearby, while Mu$^0$, with an unpaired electron in the vicinity of the $\mu^+$,
is referred to as being \textit{paramagnetic}. In an experiment where Mu$^{0}$ and Mu$^{D}$ coexist, 
but the frequencies of Mu$^{0}$ are too high to be resolved, the observed decay asymmetry $A_{D}$ 
will be less than the maximal value. This reduction of $A_D$ is commonly called a ``missing'' fraction.
The presence of free charge carriers usually causes depolarization of the Mu state
(i.e. a damping of $A$ as a function of time) either by electron or hole capture, or by spin-exchange
collisions with conduction electrons, as we will discuss in detail later.
Hence, investigations of the depolarization of Mu can provide detailed information regarding
the concentration and the type of free charge carriers in the sample.

\section*{Results}
The samples were nominally undoped commercial Ge (100) wafers from CrysTec GmbH (Berlin,
Germany) and MTI corporation (Richmond CA, USA), two side epi polished.
The CrysTec sample with a thickness of 0.15~mm was a mosaic of nine
$10\times 10$~mm$^2$ pieces, with a nominal resistivity $\rho = 30~\Omega$cm
(charge carrier density $n \sim 5\times10^{13}$~cm$^{-3}$), as specified by the supplier.
Four pieces of the Ge wafers have been sputtered with Au to create a nominally 20-nm-thin Au
layer on top of the Ge wafers. The MTI sample had a thickness of 0.5~mm and a
diameter of 1$\,''$, with $\rho = 50-60$~$\Omega$cm ($n \sim 3\times 10^{13}$~cm$^{-3}$).
The samples were glued with conductive silver on a silver coated aluminum sample plate
of the cryostat. The sample size assures that more than 95\% of the
muon beam (12~mm full-width-at-half-maxium) is hitting the sample.
The light was generated with two sources (for details, see Methods):
i) 33 commercial white and blue LEDs mounted in series on a ring inside the
radiation shield of the sample cryostat with a maximum intensity of
10~mW/cm$^2$ ~\cite{prokscha_low-energy_2012}, or ii), 
a commercial blue LED point source outside the vacuum chamber for higher intensities
up to 80~mW/cm$^2$. For the white LEDs we measured a light transmission of 40 - 50\% through
the 20-nm Au film. For this measurement we used a Au-coated transparent ZnO wafer which
was prepared simultaneously with the Ge samples. In the spectral range of 440 -- 600~nm with
about 70\% of the total luminous intensity, one can estimate the transmission to be in the range
of 28--39\%\cite{johnson_optical-constants_1972} with a peak at 500~nm. The higher measured
transmission indicates that the real thickness of the Au film is $\sim$~15~nm.

% \begin{figure*}%%[ht]
% \begin{center}
% % \includegraphics[width=0.45\linewidth]{Ge_1kG_LED-off_220K_RUN}%
% % \includegraphics[width=0.45\linewidth]{Ge_1kG_LED-off_220K_rateFast}
% % \includegraphics[width=0.45\linewidth]{Ge_1kG_LED-off_T-scan-2nd_rateFast}%
% % \includegraphics[width=0.45\linewidth]{Ge_1kG_LED-off_T-scan-2nd_asySlow}
% \end{center}
% \caption{\textbf{Germanium (100), 0.1~T applied field, 14.1 keV, mean depth $<$z$>$ = 85 nm. a,}}
% \label{fig3}
% \end{figure*}
%
% Ge results in Fig.~\ref{fig2}: at 220~K the ratio of fast/slow signal is $\sim$~1.6, no effect
% on amplitudes as a function of light intensity has been observed, in agreement with recent bulk
% data \cite{fan_influence_2008}. In bulk diamagnetic states A(B) are forming at T $>$ 200(150) K. Here, we
% do observe both states also at T $<$ 100~K with a total fraction of 25\% and fast/slow $\sim$~2.
% The diamagnetic fraction is significantly
% larger than in the bulk where it is $\simeq$~10\%. This is mainly due to a
% reduction of the Mu$_{\rm T}$ formation probability in the near-surface region  which could originate from an inhomogeneous sample with different crystal quality and/or impurity concentration near
% the surface \cite{prokscha_near-surface_2009}.

The $\mu$SR signal in Ge at a mean depth of 85~nm 
% (14.1~keV implantation energy $E_{i}$) in an applied field of 0.1~T 
is shown in Fig.~2a. The asymmetry spectrum at 220~K after cooling from room temperature
without illumination shows precession of the weakly depolarizing \textit{diamagnetic} Mu signal with
amplitude $A_D$.
% The maximum observable decay asymmetry of the LEM spectrometer at 0.1~T and $E_{i} =$~14.1~keV is $\sim$0.165.
% The observed \textit{diamagnetic} asymmetry $A_D$ of 0.09 reflects a missing fraction due to the
% presence of paramagnetic Mu$^0$ which is expected to have a fraction of 45\% at 220~K
% \cite{patterson_muonium_1988,lichti1999,fan_influence_2008,prokscha_near-surface_2009},
% in agreement with the present data. In un- and low-doped Ge Mu$^-$ is the prevailing diamagnetic charge state,
% whereas Mu$^+$ contributes only by 20\% \cite{fan_influence_2008}.
Under illumination the precession signals exhibit two components at 220~K (Fig.~2b): a fast depolarizing with 
$\sim$70\% of $A_D$, and a slow component with a fraction of $\sim$30\%. The asymmetry spectra
are well fitted\cite{suter_musrfit_2012} with the following functions:
\begin{widetext}
\begin{eqnarray}
 A_D(t) & = &  A_{\rm s} \exp(-\Lambda_{\rm s} t)\cdot \cos(\omega_{\mu} t + \phi)~(\rm{without~light,~and~at~T > 280~K}),\\
 A_D(t) & = & [A_{\rm T^{-}} \exp(-\Lambda_{\rm f} t) + A_{\rm BC^{+}} \exp(-\Lambda_{\rm s} t)]\cdot \cos(\omega_{\mu} t + \phi)~
 \nonumber \\ & & (\rm{during~and~after~illumination}),
\end{eqnarray}
\end{widetext}
% (A_{\rm T^{-}} + A_{\rm BC^{+}})
where $A_{\rm T^{-}}$ is the asymmetry of the of muons in the ${\rm Mu_T}^-$ state at the tetrahedral
interstitial site, and $A_{\rm BC^{+}}$ is the asymmetry of muons in the ${\rm Mu_{BC}^+}$ state at a bond center
site between two Ge host atoms: the T and the BC sites are the two crystallographic positions of Mu in
Ge\cite{lichti1999,fan_influence_2008}. Without light and at T~$>$~280~K, both states are indistinguishable 
and have nearly the same ``slow'' depolarization rate $\Lambda_{\rm s}$, and the ``slow'' asymmetry
$A_{\rm s} = (A_{\rm T^{-}} + A_{\rm BC^{+}})$ is the sum of the asymmetries at the two different sites. 
$\phi$ is the angle of the muon spin at $t = 0$ with respect to the positron detector.
Under illumination the appearing
``fast'' component with depolarization rate $\Lambda_{\rm f}$ is attributed to reactions of photo-generated
holes with the ${\rm Mu_T}^-$ state.
 % The observed fractions of $A_f$ and $A_s$ suggest that the fast component
% corresponds to Mu$^-$, whereas the slow component is due to Mu$^+$ \cite{fan_influence_2008}.
%It is then the Mu$^-$ state which undergoes significant interaction with photo-generated charge carriers.
By comparison with previous bulk $\mu$SR studies of photoexcited Ge\cite{fan_influence_2008}, which show 
analogous signals, the two components of equation~(2) are believed to be a consequence of the following two reactions:
i) ${\rm Mu_{BC}}^+ + e^- \rightleftharpoons {\rm Mu_{BC}^0}$ and ii),
${\rm Mu_T}^- + h^+ \rightleftharpoons {\rm Mu_T^0}$, where electron ($e^-$) or hole ($h^+$) capture,
respectively,
lead to neutralization of the corresponding diamgnetic state, i.e. conversion to the paramagnetic
neutral state. Thermal activation is responsible for the reverse reactions.
% The subscripts BC and T denote the two different crystallographic sites of the corresponding
% Mu states: BC is the bond centre between two Ge host atoms, and T is the tetrahedral interstitial site
% \cite{lichti1999,fan_influence_2008}.
For i) the reverse reaction is expected to be much faster at 220~K (thousands of MHz) 
than the carrier capture rate and the hyperfine frequency of Mu$_{\rm BC}^0$ ($\sim$~100~MHz), see Methods.
These rates imply that the loss of muon spin coherence is negligible while the muon is in the 
paramagnetic state, resulting in a slow depolarization of the Mu$_{\rm BC}^+$ signal ($A_{\rm BC^+}$).
For ii), on the contrary, the ionization rate at 220~K is of the order of the hyperfine frequency of
Mu$_{\rm T}^0$ (2360 MHz). In this case, hole capture of ${\rm Mu}^-$ is competing with thermal activation,
% at comparable transition rates. This means that the muon spends a significant amount of time as both ${\rm Mu}^-$ and
% ${\rm Mu}^0$ 
causing a significant loss of coherence in the ${\rm Mu}^-$ precession signal ($A_{\rm T^-}$), see Methods.
Furthermore, the fast rate $\Lambda_{\rm f}$ can be used as a measure of
the free hole concentration $p$: $\Lambda_{\rm f} \propto \Lambda_c^h = p\cdot v_p\cdot \sigma_c^h$~ \cite{patterson_muonium_1988}, 
where $\Lambda_c^h$ is the hole capture rate, $v_p$ is the
hole velocity, and $\sigma_c^h$ is the hole capture cross section of ${\rm Mu}^-$.
% This manifests} as
% a fast depolarizing component where $\lambda_f$ is proportional to the hole concentration.
% \textbf{Very small effect  of illumination on slow state (white LEDs: increase from 0.05 to 0.1/us at
% 220K): electrons quickly disappear by trapping at the surface: no, thermal ionization of MuBC0 (EA=170meV, 145meV
% according to PRL 2008, lichti_hydrogen_2008) is
% much faster than e- capture rate at 220 K: 14000 MHz $<->$ ~10 MHz. For MuT thermal ionization rate
% is much lower, about 160 MHz at 220 K (EA=230 meV). One might argue that the fast drop of $\lambda_f$ around
% 270 K is due to faster and faster thermal ionization; however, on subsequent cooling one should get back
% the fast component which isn't the case; this means that charge carriers must have undergo recombination
% above 270 K. For Mu_T we have to consider that the MuT precession frequencies in 1kG are
% nu12=737, nu34=1620 MHz, 88% amplitude; this means that ionization rate has to be > 1620 MHz to get back
% the full diamagnetic precession.}

The behavior of $\Lambda_{\rm f} \propto p$ and $\Lambda_{\rm s}$ as a function of temperature, illumination, and history, 
shown in Fig.~2c, constitutes the main findings of this work. On lowering the
temperature from 290~K without illumination the depolarization rate $\Lambda_{\rm s}$ slowly decreases,
as is typically observed for low doped n-type Ge \cite{patterson_muonium_1988,fan_influence_2008}. 
In this case the asymmetry spectra are fit by a single ``slow'' component, where $\Lambda_{\rm s}$
is caused by cyclic charge state transitions involving conduction electrons (but not holes)\cite{lichti1999}, 
and by nuclear magnetic dipolar fields of the $^{73}$Ge isotope. 
Turning on the light at 220~K leads to the appearance of the ``fast'' component.
Illumination with white light of intensity 10~mW/cm$^2$ for about 5~h 
saturates the depolarization rate $\Lambda_{\rm f}$ to a value of $\sim 17~\mu$s$^{-1}$.
Using blue light ($\lambda = 470$~nm) at the same intensity the saturation occurs earlier (after 2~h).
The saturation value of $\Lambda_{\rm f}$, which is proportional to the photo-induced hole concentration,
can be further increased by using shorter wavelength
light of 405~nm and increased intensity of 80~mW/cm$^2$, where saturation is already obtained after
35~min. From $\Lambda_{\rm f} \sim 40~\mu$s$^{-1}$ we estimate a hole concentration of 
$\sim 1.6\times10^{14}$/cm$^3$, see Methods.
% $10^{14} - 10^{15}$/cm$^3$\cite{fan_influence_2008}.
The most intriguing result is the persistence of the ``fast'' component.
For example, after the light is turned off at 220~K for $\sim 12$~h, $\Lambda_{\rm f}$ decreases only
to 50\% of its value directly after illumination, see Fig.~3a. The persistence can be explained by
trapping of photo-generated electrons in empty surface states (see Discussion): the negative charging of 
the surface inverts the surface region from n-type to p-type through accumulation of excess free holes.
It is only after warming the samples to 280~K that the fast component quickly disappears, and only a
single ``slow'' component ($\Lambda_{\rm s}$) remains. This indicates that a thermally activated process
liberates the electrons from the surface states, thus leading to recombination with the free holes and a
``resetting'' of the sample to its original state. In the temperature range between 270~K and
280~K $\Lambda_{\rm f}$, i.e. the hole carrier concentration, quickly disappears.
An Arrhenius plot of $\Lambda_{\rm f}$ in this temperature range allows to estimate an activation
energy of $1.1\pm 0.3$~eV for this process, see Fig.~3b. The error is the standard deviation of the fitted value of
the activation energy. The much slower decrease of $\Lambda_{\rm f}$ between
220~K and 270~K is caused by the thermally activated increase of the ${\rm Mu_T}^-$ formation rate (ionization
of ${\rm Mu_T^0}$), so that the muon spends less and less time in 
the depolarizing  ${\rm Mu_T^0}$ state when the temperature is raised, see Methods.

The whole cycle described above is reproducible.
This photo-induced persistent change of charge carrier concentration is very well reproduced in
complementary resistance measurements, see Fig.~4.
We also measured $\Lambda_{\rm f}$ as a function of implantation energy between
4 and 25~keV, corresponding to mean/maximum depths of 25/50~nm and 155/210~nm, and found no depth dependence.
This indicates that the hole concentration does not change at least
up to a depth of $\sim 200$~nm. This can be explained by the expected thickness of the
accumulation layer for a carrier concentration of $\sim 10^{14}$/cm$^3$, which is in the order
of one micrometer, see Methods.
In bulk $\mu$SR studies on Ge and Si no persistent photo-induced effects 
have been reported \cite{fan_influence_2008,fan_optically_2008,kadono1997,kadono1994}.
\section*{Discussion}
An explanation for the trapping of photo-generated electrons at the surface is provided by a
model which assumes the presence of surface acceptor states located close to the
valence band edge in Ge. These states - dangling bonds or other defects - can be easily filled, building a negative 
surface charge which drives the surface to inversion\cite{tsipas_germanium_2009,dimoulas_germanium_2009}. 
The ease of inversion appears to be
an intrinsic property of Ge surfaces, as confirmed by our present observation of inversion
in wafers from different suppliers. The model is further supported by our investigations of
n-type samples with higher doping levels of $n \sim 4\times 10^{14}$/cm$^3$ and
$n \sim 5\times 10^{17}$/cm$^3$. Whereas the first sample shows persistent inversion,
no light-induced effects are observed for the latter.
This appears to be reasonable for the sample with $n \sim 5\times 10^{17}$/cm$^3$ since the free electron 
concentration by doping is at least two orders of magnitude larger than the photo-induced carrier concentration,
resulting in a rapid recombination of the photo-generated holes. On the other hand, one can expect that
photo-generated electrons are still trapped in the surface acceptor states, so that the negative
surface charge still could generate an accumulation of holes and a depletion of electrons at the surface.
According to the model of Refs.~\onlinecite{tsipas_germanium_2009,dimoulas_germanium_2009}
this is inhibited for doping levels of $n \gtrsim 10^{16}$/cm$^3$, in agreement with our data.
% Note that the model of Ref.~\onlinecite{dimoulas_germanium_2009} predicts that inversion should be
% inhibited at $n \gtrsim 10^{16}$/cm$^3$, in agreement with our data.
It can be explained by the position of the Fermi level at the surface: for 
$n \gtrsim 10^{16}$/cm$^3$ the level is above the surface acceptor states causing these states to
be filled with electrons.
% At lower doping concentration the Fermi level at the surface $E_F^S$ is pinned close to the charge neutrality
% level (CNL) at the surface, leaving surface acceptor states unoccupied. Photo-generated electrons can
% fill these states and drive the surface to inversion. For $n \gtrsim 10^{16}$/cm$^3$
% the separation between $E_F^S$ and CNL increases and the unoccupied surface acceptor states are
% filled with electrons. 
Therefore, photo-induced electrons can't be trapped anymore at the surface in this case.
Photo-induced inversion is also destroyed by building a Schottky-barrier at the surface. After
sputtering a 20-nm-thin Au layer on top of the Ge samples -- thin enough to enable 40 - 50\% of
white light to penetrate the Au film -- we did not observe any photo-induced change of the depolarization
rate: the data are well fitted by a single ``slow'' component, equation (1), and the depolarization
rates $\Lambda_{\rm s}$ agree with the ``no light'' data in Fig.~2c. This means that photo-induced
hole accumulation at the Au/Ge interface is completely suppressed.
Strong band bending at the interface probably causes a fast separation/depletion
of photo-generated charge carriers from the region where the muons are stopping, inhibiting
hole accumulation at the interface. This further supports the interpretation that
charge accumulation at the Ge surface is driving the inversion.

In summary we have shown that a semiconductor surface can be persistently inverted
by photo-excitation, very similar to charge accumulation in a near-surface layer in
electric field transistors. It is conceivable that  -- by proper tailoring of the surface --
charge carrier concentrations can be manipulated over a larger range than presented
in this work, which is potentially interesting for future applications in opto-electronic
devices.

\section*{Methods}
\textbf{Used light sources.}
For the 33 commercial LEDs mounted in series on a ring inside the radiation shield two types of LEDs were used:
white (Avago HLMP-CW36-UX00, peak wavelengths at 460 nm and 560 nm) and blue
(Kingbright L-7113QBC-D) with a peak wavelength of 470~nm and spectral line full-width-at-half-maximum (FWHM)
of 50~nm. The maximum current/voltage  rating at room temperature is 30~mA/3.3~V.
The viewing angles are $2\Theta_{1/2} = 30^\circ$ for the white, and
$2\Theta_{1/2} = 16^\circ$ for the blue LEDs,
where $\Theta_{1/2}$ is the angle with respect to the optical center line where the luminous intensity
is $1/2$ of the center line value, i.e. the light is emitted in forward direction in a narrow cone.
The maximum intensity at the sample for this setup is 10~mW/cm$^2$. For higher intensities up to
80~mW/cm$^2$ a LED point source with four LEDs (bluepoint LED by H\"onle UV technology,
405~nm peak wave length, spectral width 9~nm FWHM) is installed
1.9~m upstream of the sample outside the vacuum chamber. The light is focussed by two optical
lenses through a vacuum viewport onto the sample.

\textbf{Ionization, charge capture and carrier concentrations.}
Thermal activation or ionization of Mu impurities is usually described by
an Arrhenius process with the ionization rate
$\Lambda_i = \Lambda_0 \exp(-E_A/kT)$, where $\Lambda_0$ is the ``attempt'' frequency
and $E_A$ the activation energy \cite{lichti1999}.
For Mu$_{\rm BC}^0$ in Ge $E_A \sim 145$~meV\cite{lichti_hydrogen_2008}, and
$\Lambda_0 = 2.9\times 10^{13}$/s~ \cite{lichti1999}. It follows $\Lambda_i \sim 14000$~MHz
at 220~K. The electron capture rate $\Lambda_c^e = n v_n \sigma_c^e \sim 230$~MHz is much
smaller. Here, $\sigma_c^e \sim 2.5\times 10^{-13}$~cm$^2$ is the electron capture
cross section\cite{lichti1999}, $v_n = 2.7\times 10^7$~cm/s is the electron velocity
in Ge at 220~K\cite{Ge_carrier_velocities},
and $n = 3\times 10^{13}$ cm$^{-3}$ is the electron density in the dark. 
A necessary condition to avoid loss of spin coherence in the precession of diamagnetic Mu$_{\rm BC}^+$
-- if it forms from a paramagnetic Mu$_{\rm BC}^0$ precursor -- is that
$\Lambda_i \gg A^{\rm HF}_{\rm BC}$, where $A^{\rm HF}_{\rm BC} \sim 100$~MHz is the hyperfine coupling
of Mu$_{\rm BC}^0$\cite{patterson_muonium_1988}. In this case, the lifetime of the neutral
state is short enough to ensure that the muon spin precession due to the hyperfine field
of the electron can be neglected. Even if there is a sizeable electron capture
rate of Mu$_{\rm BC}^+$, as it is the case here, ionization of Mu$_{\rm BC}^0$ is sufficiently
fast to avoid loss of polarization, i.e. the depolarization rate of the Mu$_{\rm BC}^+$ signal is
small.

For Mu$_{\rm T}^0$ with $E_A \sim 170$~meV, the corresponding ionization rate $\Lambda_i \sim 3700$~MHz 
at 220~K is of the order of the hyperfine coupling $A_{\rm T}^{\rm HF} = 2360$~MHz of 
Mu$_{\rm T}^0$\cite{patterson_muonium_1988}. %, which is much larger than $A_{HF}^{BC}$.
Hole capture of Mu$_{\rm T}^-$ causes therefore a temporal loss of polarization which 
leads to the observed $\Lambda_{\rm f} \sim 40~\mu$s$^{-1}$ at 220~K. Using a Monte-Carlo
simulation\cite{prokscha_monte-carlo-simulation_2012} of cyclic charge state
transitions between Mu$_{\rm T}^0$ and Mu$_{\rm T}^-$ states we find that a hole
capture rate of $\Lambda_c^h \sim 50$~MHz at 220~K is required to obtain the experimentally observed 
$\Lambda_{\rm f}$, i.e. $\Lambda_{\rm f}({\rm 220K)} \simeq 0.8\cdot\Lambda_c^h({\rm 220K})$ with
$\Lambda_c^h = p\cdot v_p\cdot \sigma_c^h$. Knowing $\sigma_c^h$ we can estimate the photo-induced
hole concentration $p$ by using the value of $\Lambda_c^h$ from the simulation and the known
hole velocity $v_p$ in Ge\cite{Ge_carrier_velocities}. We derived the hole capture cross section $\sigma_c^h$
from bulk $\mu$SR data (measured at the GPS instrument at PSI, indicated in the following by the superscript $b$) 
of a p-type Ge sample with $p^b \simeq 10^{15}$~cm$^{-3}$. In this experiment we measured $\Lambda_{\rm f}^b$(T)
between 250~K and 290~K. The observed $\Lambda_{\rm f}^b$(T) can be well reproduced by the simulation assuming
$\Lambda_c^{h,b} \simeq 300$~MHz at 220~K. We obtain $\sigma_h^c = \Lambda_c^{h,b}/(p^b\cdot v_p) 
\simeq 2\times 10^{-14}$~cm$^2$, where we used $v_p = 1.6\times 10^7$~cm/s at 220~K\cite{Ge_carrier_velocities}.
Using this value of $\sigma_h^c$ we can estimate the photo-induced hole concentration to 
$p = \Lambda_c^h/(\sigma_c^h\cdot v_p) \simeq 1.6\times 10^{14}$/cm$^3$.

The simulation also shows that $\Lambda_{\rm f}$ is decreasing as a function of temperature because of the exponentially 
increasing $\Lambda_i$, in good agreement with the observed decrease of $\Lambda_{\rm f}$ between 220~K and 265~K. 
Note, that $\Lambda_c^h$ is also expected to increase with temperatur T, because of 
$v_p \propto \sqrt{\rm T}$\cite{Ge_carrier_velocities}. An increase of $\Lambda_c^h$ should cause an \textit{increase}
of $\Lambda_{\rm f}$. However, $\Lambda_i$ changes exponentially with T and dominates the dynamics of the 
Mu$_{\rm T}^0$ $\rightleftharpoons$ Mu$_{\rm T}^-$ charge cycles. 
The much faster decrease of $\Lambda_{\rm f}$ between 270~K and 280~K can not be explained by this model which assumes
charge cycles due to the presence of holes. The disappearance of
holes in this temperature range -- by recombination with electrons liberated from the surface states -- 
stops the Mu$_{\rm T}^0$ $\rightleftharpoons$ Mu$_{\rm T}^-$ charge cycles, which explains the fast drop of 
$\Lambda_{\rm f}$. 

The photo-induced hole carrier concentration can be independently estimated by the change of resistance $R$
after illumination. The measurements in Fig.~4 yield a persistent reduction of $R$ by a factor 2-3.
The initial resistivity $\rho$ at room temperature of 50-60~$\Omega$cm corresponds to 
$n \simeq 3\times 10^{13}$/cm$^3$ ~\cite{cuttriss_relation_1961}. 
After illumination, $\rho \sim$~20-25~$\Omega$cm, and assuming that
the conductivity is now p-type we can estimate the hole carrier concentration to be 
$p \simeq 1.3-1.6\times 10^{14}$/cm$^3$ ~\cite{cuttriss_relation_1961}, in excellent agreement with the
$\mu$SR data.

\textbf{Thickness of hole accumulation layer.}
The thickness $d$ of the hole accumulation layer can be estimated by
$d = \sqrt{2\varepsilon_r\varepsilon_0 U/(e p)}$ ~\cite{sze2007}, where $\varepsilon_r = 16$ is the dielectric constant of
Ge, $\varepsilon_0$ is the permittivity of free space, $U$ is the surface potential, $e$ the electron charge,
and $p$ the hole concentration. Assuming that the surface potential $U$ is close to the charge neutrality
level at the Ge surface\cite{dimoulas_germanium_2009} we use $U  \sim 0.3$~V to obtain $d = 1.8$~$\mu$m for 
$p = 1.6\times 10^{14}$/cm$^3$.

% The measurement of the $\mu$SR relaxation rate allows therefore conclusions
% on the charge carrier concentration. Thermally activated ionization of
% Mu defect states located in the band gap changes the amplitudes of the
% charged and neutral Mu states. In the diamond-like crystal lattices of
% Si and Ge two stopping sites of the $\mu^+$ are known. A hydrogen-like donor
% state at the bond-center between two host atoms with anisotropic hyperfine
% interacrtion, which is either positive

% \bibliographystyle{prsty}
% \bibliography{tp_semiconductor}

\section*{Acknowledgements}
We thank M. Horisberger for sputtering the Au films, and H.P. Weber for
his excellent technical support.

\begin{figure*}[t]
\begin{center}
\includegraphics[width=0.95\linewidth]{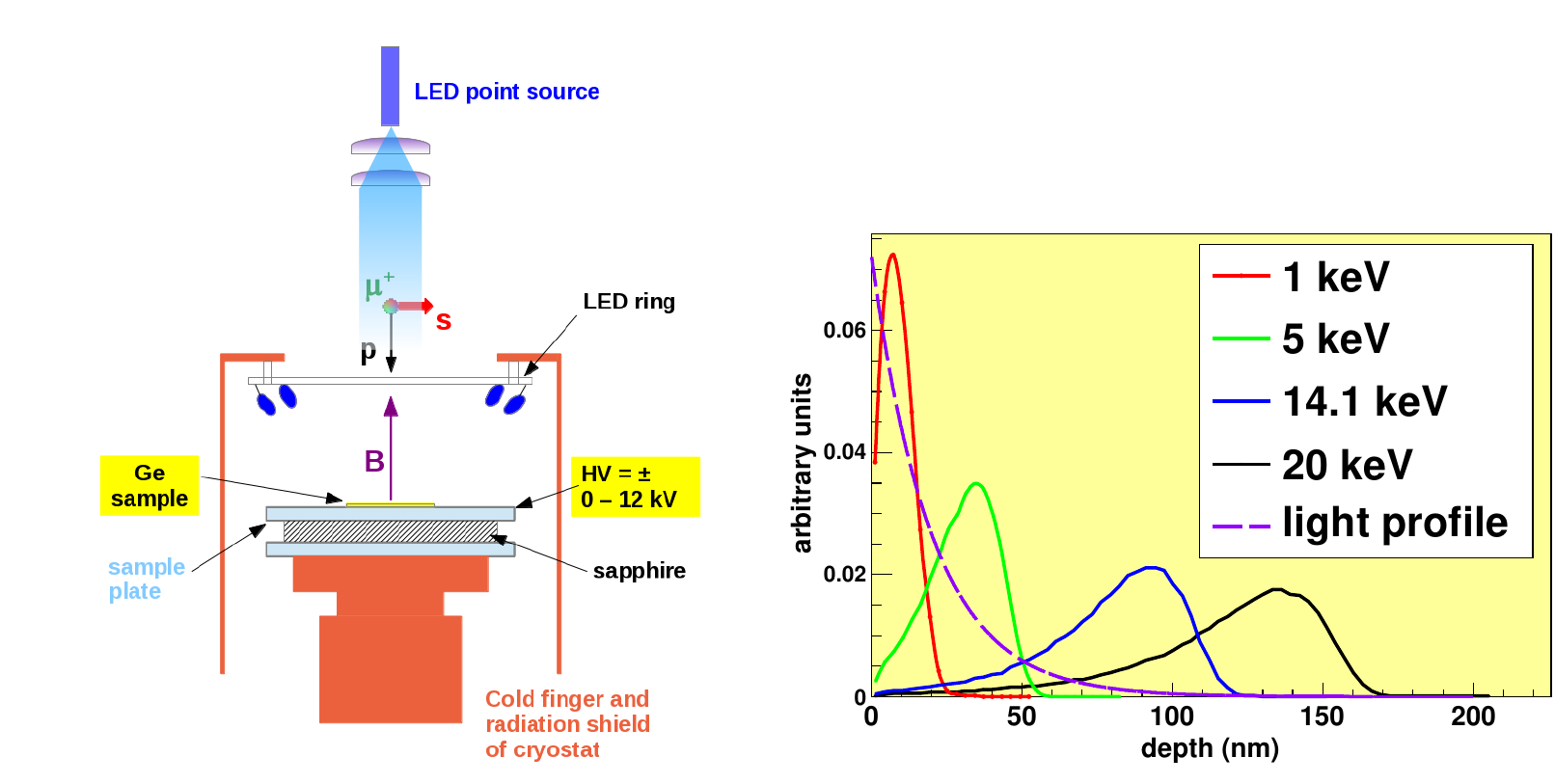}
\end{center}
\caption{\textbf{Schematic of experimental setup and muon stopping and light absorption profiles in germanium.}
Muon stopping profiles at the indicated implantation energies were calculated with the Monte-Carlo code 
{\ttfamily TrimSP}\cite{trimsp,morenzoni_implantation_2002} from 1 keV to 20 keV.
A light absorption coefficient $\alpha \sim 5\cdot 10^{5}/cm$ has been assumed
for the used blue light sources with peak wavelengths of 470~nm and 405~nm
\cite{philipp_optical_1959}, corresponding to photon energies of 2.6 and 3.0~eV, respectively.
The pressure at the sample was $\leq 10^{-8}$~mbar.}
\label{fig1}
\end{figure*}

\begin{figure*}[t]
\begin{center}
 \includegraphics[width=0.95\linewidth]{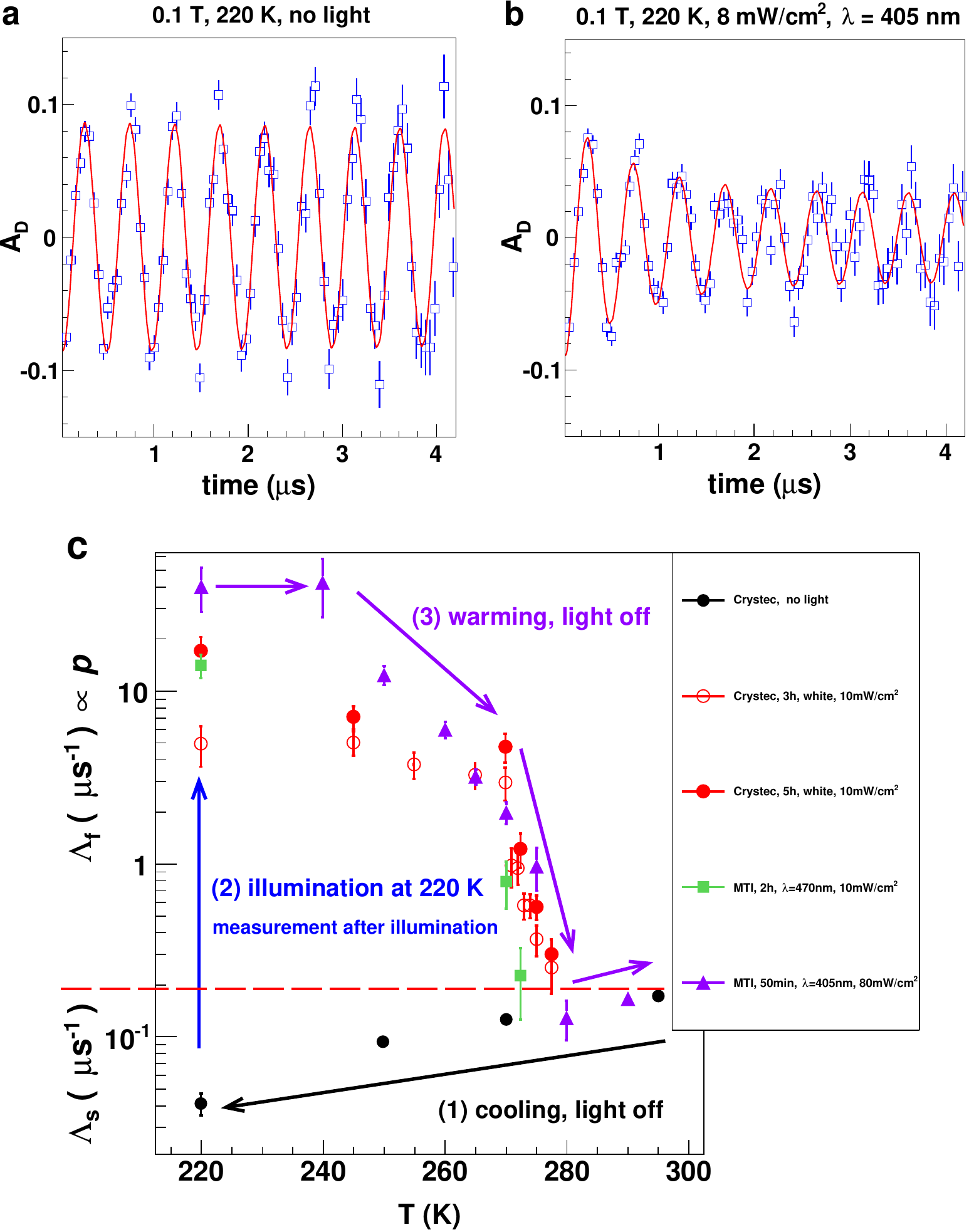}
\end{center}
\vspace{-0.7cm}
\caption{\textbf{Germanium (100), 0.1~T applied field, implantation energy 14.1 keV, mean depth
 $<$z$>$ = 85 nm. a}, $\mu$SR asymmetry $\rm A_D$ at 220~K, in the dark, MTI sample. For better
 illustration data are shown in a rotating reference frame (rrf) of 11.5~MHz.
 \textbf{b}, $\rm A_D$ after one hour illumination
 with blue LEDs ($\lambda = $ 405~nm, $\sim 8~$mW/cm$^2$), rrf representation. A fast depolarizing component is
 visible in the first 3~$\mu$s, and a much slower component at later times. Solid lines are fits of equation (1) to
 the data. \textbf{c}, The depolarization rate $\Lambda_{\rm f}$ of the fast component as a function of temperature T
 and history for two
 different commercial
 wafers (CrysTec GmbH and MTI corporation): samples are cooled in the dark to 220~K (1), followed by illumination
 at 220~K for a period indicated in the legend (2), and measured after turning the light off (3). The fast component
 persists indicating the presence of free holes. After warming above 280~K the fast component disappears.
 The time of the warming sequence (light off)
 varied between 4~h (closed squares), 9~h (closed circles and triangles) and 19~h (open circles).
 }
\label{fig2}
\end{figure*}

\begin{figure*}[t]
%\vspace{1cm}
\begin{center}
\includegraphics[width=0.95\linewidth]{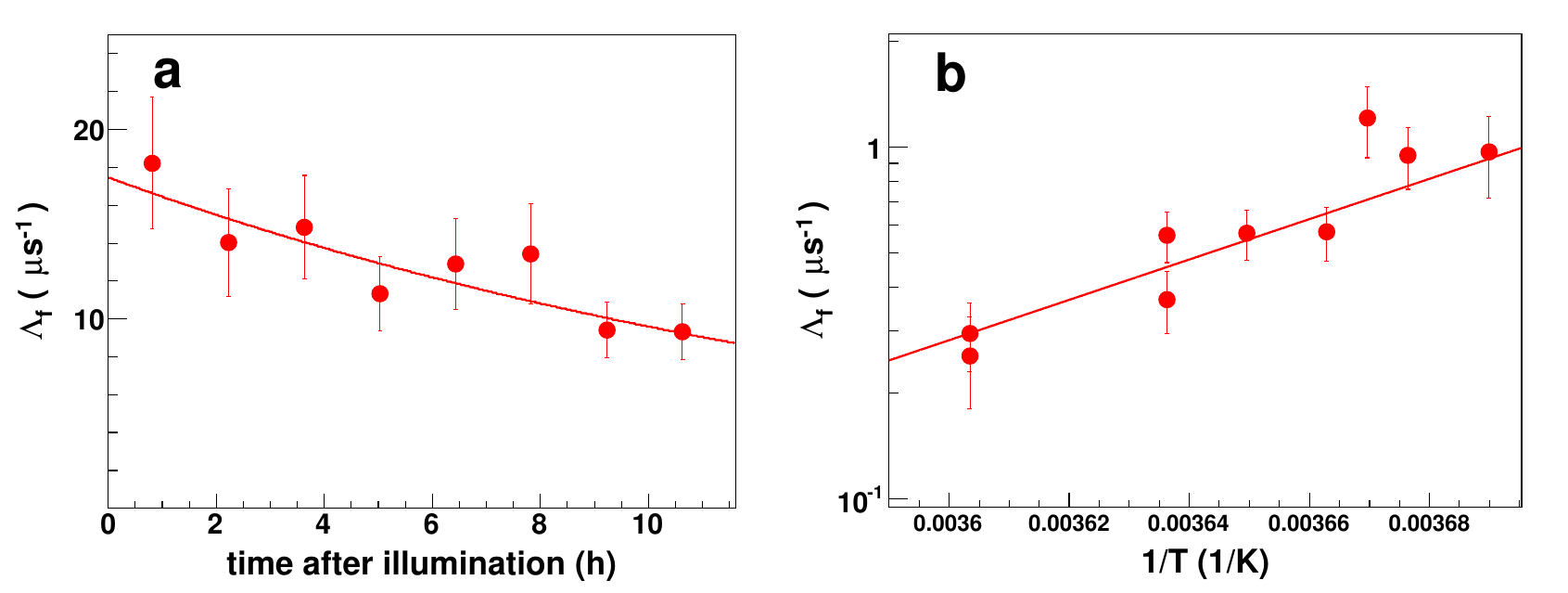}
\end{center}
\caption{\textbf{Germanium (100), fast depolarization rate  $\Lambda_{\rm f}$, CrysTec sample.} 
 \textbf{a}, $\Lambda_{\rm f}$ at temperature T = 220~K as a function of time after terminating the illumination with 
 white LEDs 
 at 10 mW/cm$^2$ for 5~h. The solid line is a fit of the exponential function $\exp(-t/\tau_D)$, where $t$ is the time
 and $\tau_D$ the time constant of the decrease of $\Lambda_{\rm f}$. The fit yields $\tau_D = 16.7(5.5)$~h, which 
 corresponds to a half-life of about 12~h. The error indicates the standard deviation of the fit result. \textbf{b}, 
 Arrhenius plot of $\Lambda_{\rm f}$ between 270~K and 280~K.
 Fit function is $\Lambda_{\rm f}({\rm T}) = C \exp(E_{A}/k_B T)$, where $C$ is a constant, $E_A = 1.1(3)$~eV is the 
 activation energy, and $k_B$ is the Boltzmann constant. 
 }
% \label{fig3}
\end{figure*}

\begin{figure*}[t]
\begin{center}
\includegraphics[width=0.95\linewidth]{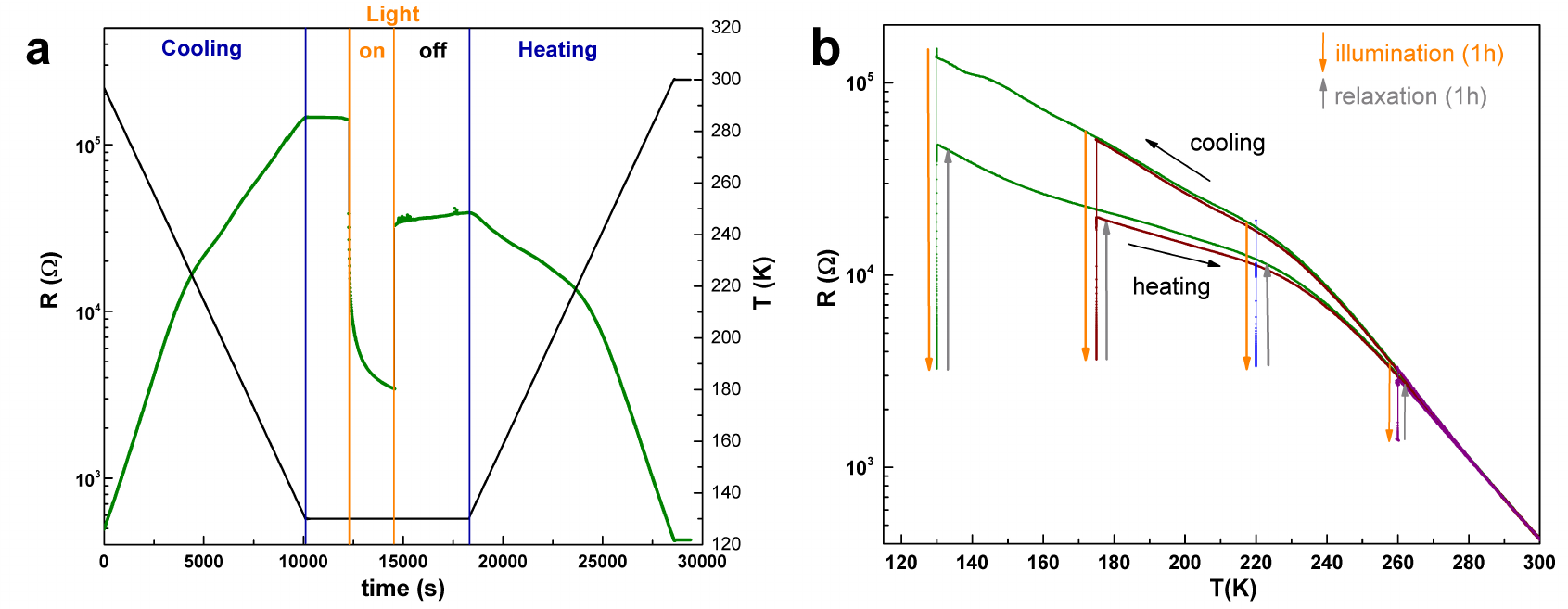}
\end{center}
\caption{\textbf{Germanium (100), MTI sample, resistance measurements. a},
resistance R as a function of time during cooling from 300~K to 130~K. Light
is turned on at $\sim$~12200~s, $\lambda = $~405~nm, 80~mW/cm$^2$. The drop of resistance is
caused by the photo-generated charge carriers. After switching off the light at $\sim$~14600~s
the resistance doesn't return to its initial value R$_0$, but only to $\sim$~0.5~R$_0$, which
is clear evidence for a persistent charge carrier accumulation in the sample. The slow
increase of R as a function of time indicates the slow recombination of electrons trapped
at the surface with accumulated holes close to the surface. At $\sim$~18300~s a slow warmup of 
the sample to 300~K is initiated. \textbf{b}, R as a function of temperature T. The sample
was cooled in the dark to various temperatures and illuminated for 1h at that temperature, 
followed by a wait (``relaxation'') for 1h in the dark, and a slow warmup to 300~K. The
cooling and warming-after-illumination curves merge between 260 and 280~K, which matches
exactly the temperature range where the $\mu$SR data show the thermally activated recombination of 
surface electrons with accumulated holes near the surface.}
% \label{fig4}
\end{figure*}

\end{document}